\documentstyle[aps,preprint]{revtex}
\begin{document}
\title{
Gauge Symmetry Enhancement and Radiatively Induced Mass 
in the Large N Nonlinear Sigma Model
}

\author{
Taichi Itoh$^{1}$\footnote{Email address: taichi@knu.ac.kr},
Phillial Oh$^2$\footnote{Email address: ploh@dirac.skku.ac.kr},
and Cheol Ryou$^2$\footnote{Email address: cheol@newton.skku.ac.kr}
}
\address{
$^1$Department of Physics, Kyungpook National University, 
Taegu 702-701, Korea\\
$^2$Department of Physics and Institute of Basic Science, 
Sungkyunkwan University, Suwon 440-746, Korea
}


\maketitle
\draft

\begin{abstract}
We consider a hybrid of nonlinear sigma models in which two complex 
projective spaces are coupled with each other under a duality. 
We study the large N effective action in 1+1 dimensions. 
We find that some of the dynamically generated gauge bosons acquire 
radiatively induced masses which, however, vanish along the self-dual 
points where the two couplings characterizing each complex projective 
space coincide. These points correspond to the target space of the 
Grassmann manifold along which the gauge symmetry is enhanced, 
and the theory favors the non-Abelian ultraviolet fixed point.
\end{abstract}

\vspace{5mm}



The nonlinear sigma models (NLSM's) in which the dynamical fields take 
values in some target manifolds have been a subject of extensive research in
theoretical physics due to their wide range of physical applications and their 
relevance with geometrical aspects of quantum theory \cite{band,zakr}. 
Especially, the large-$N$ analysis \cite{poly} of this model has proved to
exhibit many remarkable physical properties, such as dynamical generation of
gauge bosons, nonperturbative renormalizability, dimensional
transmutation, and phase transitions in the lower dimensional space-time
\cite{bard,aref,itoh}.

One of the well studied NLSM is the complex projective $CP(N)$ model 
\cite{golo} where the target space is given by the complex projective space
$CP(N)\equiv SU(N)/SU(N-1) \times U(1)$. The purpose of this paper is to
investigate the large-$N$ limit of the NLSM for some other target space
and to re-examine the issue of the dynamical generation of non-Abelian gauge
bosons in 1+1 dimensions. Especially, we first study the specific
target space given by the Grassmann coset space $Gr(N,2)\equiv SU(N)/SU(N-2)
\times U(2)$ \cite{pisa}.
It turns out that this NLSM can be written as a hybrid of two $CP(N)$ models 
coupled to each other with the same coupling constant for each complex
projective space and the interaction terms respect the dual exchange symmetry
between the two sectors [see Eq.~(\ref{lag2})]. We observe that there exists 
a manifest dual symmetry between the two sectors, and the Grassmann manifold
corresponds to a self-dual case with equal coupling constants. 
If we start from different coupling constants for each complex
space for the generality, this leads to the target space belonging to the
so-called flag manifold \cite{helg} ${\cal M}=SU(N)/SU(N-2)\times U(1)\times
U(1)$. The dynamically generated gauge bosons would have $U(1)\times U(1)$
gauge symmetry. These observations lead to our main motivation for this
work, that is, a study of self-duality in the coupling constant space and
subsequent enhancement of gauge symmetry. In order to investigate this
issue, we analyze the large-$N$ mass gap equations, and renormalization group 
(RG) properties in the coupling constant spaces. We also explicitly compute 
the large-$N$ effective action. We find that some of the dynamically
generated gauge bosons acquire radiatively induced finite mass terms and
gauge noninvariant interaction away from the self-dual points, leading to a
local $U(1)\times U(1)$ symmetry. However, they vanish at the self-dual points
enhancing the gauge symmetry to a non-Abelian $U(2)$ symmetry. 
The ultraviolet (UV) fixed point corresponds to a special self-dual point 
and the theory prefers the non-Abelian phase in the UV limit. 
Even though the dynamical generation of non-Abelian gauge bosons for the
Grassmann target space has been discussed before \cite{band,duer}, 
the way in which the enhancement of gauge symmetry at the fixed point occurs 
through the RG evolution has not been addressed so far.

Let us consider a Lagrangian which is given by 
\begin{eqnarray}
{\cal L}_{0} &=& \frac{1}{g_1^2}
|(\partial_\mu +iA_\mu)\psi_1|^2 +\frac{1}{g_2^2}|(\partial_\mu
+iB_\mu)\psi_2|^2 +\frac{1}{4}
\left(\frac{g_1}{g_2}+\frac{g_2}{g_1}\right)C_\mu^* C^\mu 
\nonumber \\ &&
-i\frac{1}{\sqrt{g_1g_2}}C_\mu^* \psi_1^\dagger \partial^\mu \psi_2
-i\frac{1}{\sqrt{g_1g_2}}C_\mu \psi_2^\dagger \partial^\mu \psi_1, 
\label{lag2}
\end{eqnarray}
where $\psi_1$ and $\psi_2$ are two orthonormal complex $N$ vectors
such that $\psi^\dagger_i\psi_j=\delta_{ij}~(i,j=1,2)$. 
The above Lagrangian describes two $CP(N)$ models 
each described by $\psi_1$ and $\psi_2$ with
coupling constants $g_1$ and $g_2$, respectively, 
coupled through derivative coupling.
There is a manifest dual symmetry between sectors 1 and 2,
$A_\mu$ and $B_\mu$, and $C_\mu$ and $C^*_\mu$.
When $g_1=g_2$, the above model corresponds to the
nonlinear sigma model with the target space of Grassmann manifold
$Gr(N,2)=SU(N)/SU(N-2)\times U(2)$. 
Let us write the above Lagrangian in the more conventional form
in terms of the $N\times2$ matrix $Z$:
\begin{equation}
Z=\left[\psi_1, \psi_2\right],\quad\longleftrightarrow\quad 
Z^\dagger=\left[{\psi_1^\dagger \atop \psi_2^\dagger}\right].
\end{equation}
 We first introduce new sets of coupling constants defined by 
$g\equiv\sqrt{g_1g_2}$ and $r\equiv g_2 /g_1$. Then, we consider
\begin{equation}
{\cal L}=\frac{1}{g^2}{\rm tr}
\left[(D_\mu Z)^\dagger (D^\mu Z)-\lambda(Z^\dagger Z-R)\right],
\label{lag1}
\end{equation}
where we collected the orthonormal constraints into a $2\times2$ Hermitian
matrix $\lambda$ which transforms as an adjoint representation
under the local $U(2)$ transformation, and the $R$ matrix given by 
\begin{equation}
\lambda=\left[\,
{\lambda_1 \atop \lambda_3^*}\quad
{\lambda_3 \atop \lambda_2}
\right],~~~
R=\left[\,{r \atop 0}\quad{0 \atop r^{-1}}\right],
\end{equation}
with a real positive $r$ \cite{macf}.
The covariant derivative is defined as 
$D_\mu Z \equiv\partial_\mu Z -Z\tilde{A}_\mu$ with a $2\times2$ 
anti-Hermitian matrix gauge potential 
$\tilde{A}_\mu\equiv-i\tilde{A}_\mu^a T^a$ 
associated with the local $U(2)$ symmetry. 
Each component of $\tilde{A}_\mu$ is assigned as
\begin{equation}
\tilde{A}_\mu =-i\left[\,
{A_\mu \atop \frac{1}{2}C_\mu^*}\quad
{\frac{1}{2}C_\mu \atop B_\mu}
\right].
\end{equation}
The on-shell equivalence between Lagrangians Eqs.\ (\ref{lag2}) and 
(\ref{lag1}) will be discussed shortly. The Lagrangian Eq.\ (\ref{lag1}) is
invariant under the local $U(2)$ transformation for $r=1$, whereas the $R$
with $r\neq1$ explicitly breaks the $U(2)$ gauge symmetry down to $U(1)_A
\times U(1)_B$, where $U(1)_A$ and $U(1)_B$ are generated by $T^0 \pm
T^3$, respectively. Thus the symmetry of our model is 
$[SU(N)]_{\rm global} \times [U(2)]_{\rm local}$ for $r=1$, 
while $[SU(N)]_{\rm global} \times 
[U(1)_A \times U(1)_B]_{\rm local}$ for $r\neq 1$.
Invoking the hidden local symmetry \cite{bal}, we infer that the theory
with $r \neq 1$ corresponds to NLSM on the flag manifold
${\cal M}=SU(N)/SU(N-2)\times U(1)\times U(1)$.

In order to carry out the path integration in the large-$N$ limit, we 
arrange the Lagrangian (\ref{lag1}) in terms of the $2N\times 2N$ matrix form
\begin{equation}
{\cal L}=\frac{1}{g^2}\,[\psi_1^\dagger,\psi_2^\dagger]\,(M^T \otimes I)
\left[{\psi_1 \atop \psi_2}\right]
+\frac{r}{g^2}\lambda_1 +\frac{1}{rg^2}\lambda_2,
\label{lag3}
\end{equation}
where $I$ is an $N\times N$ unit matrix and $M$ is a $2\times 2$ matrix 
operator given by
\begin{eqnarray}
M &\equiv& G^{-1}-\Gamma(\tilde{A}), \label{matrix}\\
G^{-1} &\equiv& -\Box -\lambda\,\,=\,\,
\left[\,
{-\Box -\lambda_1 \atop -\lambda_3^*}\quad
{-\lambda_3 \atop -\Box -\lambda_2}
\,\right], \label{Ginv}\\
\Gamma(\tilde{A}) &\equiv& 
-\tilde{A}_\mu \hat{\partial}^\mu
+\tilde{A}_\mu \tilde{A}^\mu, 
\label{gamma}
\end{eqnarray}
where the differential operator 
$\hat{\partial}^\mu\equiv\partial^\mu -\overleftarrow{\partial^\mu}$ 
must be regarded as not operating on the gauge potential $\tilde{A}_\mu$.
One can show that imposing the on-shell constraints
$\psi_1^\dagger\psi_1=r$, $\psi_2^\dagger\psi_2=r^{-1}$, $\psi_1^\dagger
\psi_2 =\psi_2^\dagger \psi_1 =0$, and in terms of rescaling given by
 \begin{eqnarray}
\frac{\psi_1}{g}\rightarrow
\frac{\psi_1}{g_1},~\frac{\psi_2}{g}\rightarrow \frac{\psi_2}{g_2},
~ \frac{C_\mu}{g}\rightarrow C_\mu,~\frac{C_\mu^*}{g}\rightarrow 
C_\mu^*, 
\end{eqnarray}
the Lagrangian Eq.\ (\ref{lag3}) reduces Eq.\ (\ref{lag2}).
We note that in Eq.~(\ref{lag3}), we never used the on-shell constraints 
so that the quadratic term of $C_\mu^* C^\mu$ has been absorbed into the 
matrix $M$. 


Let us focus on the two dimensions from here on. Detailed analysis in other 
dimensions will be reported elsewhere \cite{itoh1}.
The large-$N$ effective action is given by path integrating $Z$ and 
$Z^\dagger$, or equivalently $\psi_1$, $\psi_1^\dagger$, $\psi_2$, and 
$\psi_2^\dagger$. We obtain
\begin{equation}
S_{\rm eff}=\int_x {\cal L}+iN\ln{\rm Det}M. \label{eff1}
\end{equation}
According to the Coleman-Mermin-Wagner theorem, which states that 
there is no spontaneous breaking of any continuous global symmetry
in dimensions two or less, we can set the vacuum expectation values of
$\psi_1$ and $\psi_2$ equal to zero from the beginning in the effective
potential such that 
\begin{equation}
V_{\rm eff}=-\frac{1}{N\Omega}S_{\rm eff}
[\psi_{1,2}=0,\lambda_{1,2,3}=m^2_{1,2,3},
\tilde{A}_\mu=0],
\end{equation}
where $\Omega$ denotes the space-time volume.
Then we obtain the large-$N$ effective potential as
\begin{equation}
V_{\rm eff}=-\frac{m^2_1}{Ng^2}r -\frac{m^2_2}{Ng^2}r^{-1}
-i\Omega^{-1}\ln{\rm Det}\,G^{-1},
\end{equation}
from which the gap equations are schematically given as follows:
\begin{eqnarray}
\frac{\partial V_{\rm eff}}{\partial m^2_3} &=& 
-\int\! \frac{d^2 k}{(2\pi)^2}
\frac{2m^2_3}{(k^2 +m^2_+)(k^2 +m^2_-)}=0,
\label{gap3}\\
\frac{\partial V_{\rm eff}}{\partial m^2_1} &=& 
-\frac{1}{Ng^2}r+\int\! \frac{d^2 k}{(2\pi)^2}
\frac{k^2 +m^2_2}{(k^2 +m^2_+)(k^2 +m^2_-)}=0,
\label{gap4}\\
\frac{\partial V_{\rm eff}}{\partial m^2_2} &=& 
-\frac{1}{Ng^2}r^{-1}+\int\! \frac{d^2 k}{(2\pi)^2}
\frac{k^2 +m^2_1}{(k^2 +m^2_+)(k^2 +m^2_-)}=0
\label{gap5}.
\end{eqnarray}
Here the loop momenta are Euclidean
and $m^2_\pm$ are given in terms of $m^2_{1,2,3}$ by
\begin{equation}
m^2_+ +m^2_- =m^2_1 +m^2_2,\quad
m^2_+ m^2_- =m^2_1 m^2_2 -m^4_3.
\label{lambda1}
\end{equation} 
Since Eq.\ (\ref{gap3}) simply states $m_3 =0$, we can choose for example 
$m^2_+ =m^2_1$, $m^2_- =m^2_2$ after setting $m_3 =0$ in Eq.\ (\ref{lambda1}).
Then the gap equations are 
given by two decoupled sets of equations expressed by
\begin{eqnarray}
0 &=& 
\frac{1}{Ng_1^2}-\frac{1}{4\pi}\ln \frac{\Lambda^2}{m_1^2},
\label{gap14}\\
0 &=& 
\frac{1}{Ng_2^2}-\frac{1}{4\pi}\ln \frac{\Lambda^2}{m_2^2}
,\label{gap15}
\end{eqnarray}
where $\Lambda$ is the cutoff of the theory. The above equations
yield two mass scales given by
\begin{eqnarray}
m_i^2=\Lambda^2\exp\left[-\frac{4\pi }{Ng_i^2}\right]~~(i=1,2).
\label{dimtra}
\end{eqnarray}
Imposing the cutoff independence of the mass scales, 
$\Lambda dm_i/d\Lambda=0$ leads to the
Callan-Symanzik $\beta$ functions
\begin{eqnarray}
\beta_i (g_i)=\frac{dg_i}{d\ln\Lambda}=-\frac{Ng^3_i}{4\pi},
\label{rg2}
\end{eqnarray}
which show the asymptotic free behaviors of both couplings and 
a UV fixed point at the origin of the coupling constant space $(g_1,g_2)$. 
Note that when $m_1=m_2$, we have $g_1=g_2$ and the corresponding
nonlinear sigma model is defined on the Grassmann manifold.

Let us discuss the dynamical generation of gauge bosons in our model.
It has been discussed before that if we start
from the same coupling constants $r=1$ in the Lagrangian Eq.\ (\ref{lag3}),
the effective action generates non-Abelian gauge bosons with a local
$U(2)$ symmetry \cite{duer}, rendering all four gauge bosons
$A, B, C$, and $C^*$ massless. Our main objective here is to compute
the radiatively induced mass terms for the gauge bosons in the generic case
where $g_1\neq g_2$, hence $m_1\neq m_2$.
The large-$N$ effective action Eq.\ (\ref{eff1}) is schematically expanded 
such that 
\begin{equation}
S_{\rm eff}=\int_x {\cal L}+iN\ln{\rm Det}\,G^{-1}
-iN\sum_{n=1}^{\infty}\frac{1}{n}{\rm Tr}\left[G\Gamma(\tilde{A})\right]^n.
\label{eff2}
\end{equation}
The boson propagator $G$ becomes a diagonal $2\times2$ matrix due to 
the gap equation solution $m_3 =0$. We neglect the fluctuation
fields coming from $\lambda_{1,2,3}$ around $m^2_{1,2,3}$.
The diagrams which arise from  $n=1,2,3,4$
can contribute to the Yang-Mills action.
The mass term comes from $n=1$ and $n=2$. 
For $n=2$, we have three diagrams with two, three, and four point functions.
The two point vacuum polarization function provides
the gauge bosons with the kinetic terms and the mass term for $C$, $C^*$ 
fields in the case $m_1\neq m_2$. 
The contributions from both $n=1$ and $n=2$ are explicitly given by the 
integral
\begin{eqnarray} 
&& -iN\frac{1}{2}{\rm Tr}\left[G\tilde{A}_\mu \hat{\partial}^\mu
G\tilde{A}_\nu \hat{\partial}^\nu \right]
-iN{\rm Tr}\left[G\tilde{A}_\mu\tilde{A}^\mu \right]
\nonumber \\ && = 
\frac{N}{2}\sum_{ij}\int_x\,\tilde{A}^\mu_{ij}(x)
\Pi_{\mu\nu}^{ij}(i\partial_x)\tilde{A}^\nu_{ji}(x),
\label{vacuum}
\end{eqnarray}
where
\begin{equation}
\Pi_{\mu\nu}^{ij}(p) = - \int\!\frac{d^2 k}{i(2\pi)^2}
\frac{(2k+p)_\mu (2k+p)_\nu}{(k^2 -m_i^2)[(k+p)^2 -m_j^2]}
+\int\!\frac{d^2 k}{i(2\pi)^2}\frac{2g_{\mu\nu}}{k^2 -m_i^2}.
\end{equation}
This vacuum polarization can be explicitly computed to yield
\begin{equation}
\Pi_{\mu\nu}^{ij}(p)=
\left(g_{\mu\nu}-\frac{p_\mu p_\nu}{p^2}\right)\Pi_T^{ij}(p)
+\left(\frac{p_\mu p_\nu}{p^2}\right)\Pi_L^{ij}(p),
\label{mas}
\end{equation}
where the transverse function $\Pi_T$ and the longitudinal one $\Pi_L$ 
are obtained as
\begin{eqnarray}
\Pi_T^{ij}(p) &\equiv& 
\frac{1}{2\pi}\left[\ln\frac{m_j}{m_i}
-\int_0^1 dx \ln\frac{K^{ij}}{m_i^2}\right],\\
\Pi_L^{ij}(p) &\equiv& \frac{\left(m_j^2 -m_i^2\right)^2}{4\pi p^2} 
\left[\frac{2}{m_j^2 -m_i^2}\ln\frac{m_j}{m_i}
-\int_0^1 dx \frac{1}{K^{ij}}\right],
\end{eqnarray}
with $K^{ij}\equiv x m_j^2 +(1-x) m_i^2 -x(1-x)p^2$. 
Moreover we see that
\begin{eqnarray}
\Pi_T^{ij}(p) = c^{ij} + p^2 f_T^{ij} (p),~~~ 
\Pi_L^{ij}(p) = c^{ij} + p^2 f_L^{ij} (p),
\label{mass}
\end{eqnarray}
where the same constant $c^{ij}$ arises in the leading terms of 
both $\Pi_T$ and $\Pi_L$, and is given by
\begin{equation}
c^{ij} = \frac{1}{2\pi}\left[
1-\frac{m_j^2 +m_i^2}{m_j^2 -m_i^2}\ln\frac{m_j}{m_i}\right].
\end{equation}
We note that, despite its appearance, $c^{ij}$ vanishes for $m_i=m_j$. 
Then the vacuum polarization can be written as
\begin{equation}
\Pi_{\mu\nu}^{ij}(p)=c^{ij} g_{\mu\nu} 
+(p^2 g_{\mu\nu}-p_\mu p_\nu) f_T^{ij}(p)
+p_\mu p_\nu f_L^{ij}(p),
\end{equation}
where both $c^{ij}$ and $f_L^{ij}$ vanish when $i=j$ 
so as to provide the $A$ ($B$) boson with the $U(1)_A$ ($U(1)_B$) 
gauge invariant kinetic term. On the other hand, they remain nonzero when
$i\neq j$ and provide the $C$ boson with the mass given by
\begin{equation}
M_C = \sqrt{\frac{-c^{12}}{f_T^{12} (0)}}=\left|m_1^2 -m_2^2 \right|
\sqrt{\frac{(m_1^2 +m_2^2)\ln(m_1/m_2)+m_2^2 -m_1^2}
{(m_1^4 -m_2^4)/2-2 m_1^2 m_2^2 \ln(m_1/m_2)}}.
\end{equation}
This is one of the main results of our paper. We note that
the above mass does not vanish when $m_1\neq m_2$, 
which in turn implies $g_1\neq g_2$ from the mass gap equations (\ref{gap14})
and (\ref{gap15}). It is also symmetric under the exchange of $m_1$ and $m_2$. 
When $r=1$ ($m_1 =m_2$), both $c^{12}$ and $f_L^{12}$ 
become zero.
The three point function with one seagull does not contribute 
to the Yang-Mills action. 
The four-point vertex with two seagulls
also contributes to the kinetic term of Yang-Mills action with other
contributions from $n=3,4$. 

Combining the relevant diagrams up to $n=4$, 
we obtain the Yang-Mills effective action for $m_1=m_2=m$ 
given by \cite{duer} 
\begin{eqnarray}
{\cal L}_{\rm eff}=\frac{N}{48\pi m^2}\,{\rm tr}\,
F_{\mu\nu}F^{\mu\nu}(\tilde A),
\end{eqnarray}
where $F_{\mu\nu}(\tilde{A})\equiv \partial_{[\mu}\tilde{A}_{\nu]}+
[\tilde{A}_\mu, \tilde{A}_\nu]$
is the gauge field strength of $\tilde A_\mu$.
When $m_1\neq m_2$, the effective action contains interactions
that are not $U(2)$ gauge invariant. These terms and the nature of their 
interactions will be reported elsewhere \cite{itoh1}.
In passing, we observe that the large-$N$
effective action is renormalizable because the only UV divergence is the
one that arises in the gap equation and the other possible UV
divergences in the vacuum
polarization function are forbidden by the gauge symmetry.
The renormalization conditions Eqs.\ (\ref{gap14}) and (\ref{gap15}) are 
enough to realize the UV finite large-$N$ theory. 
The higher order corrections in $1/N$ expansion can be systematically 
renormalized by using the counter terms, which are provided by the large-$N$
effective action.


We have performed the large-$N$ path integral of a coupled $CP(N)$ model
with dual symmetry and have analyzed the vacuum structure and 
renormalization in 1+1 dimensions. The large-$N$ gap equation analysis
yields two decoupled gap equations whose solution ensures the
renormalizability. We also have computed the effective action, and
have found that some of the dynamically generated gauge bosons acquire 
radiatively induced finite mass terms away from the self-dual
points, and the gauge symmetry is reduced to its subgroup \cite{break}.
We note that the theory favors the conformal fixed point and the non-Abelian
phase in the ultraviolet limit. Also the classical dual symmetry is not broken
by the nonperturbative radiative corrections. 

We would like to emphasize that the mass generation of $C$ gauge bosons is a 
genuine quantum effect away from the self-dual points. 
The finite mass term is determined unambiguously and is independent of the
gauge invariant regularization scheme employed. In our scheme, the 
$C$ boson mass arises from a purely finite term of $\Pi_T$ of Eq.~(\ref{mass}).
This unambiguity is in contrast with some other radiative corrections 
in quantum field theory which are finite but undetermined
\cite{jack}.

We could extend our model to describe other types of symmetry
reduction patterns and to include supersymmetry. For example, if we envisage
the Grassmann space $Gr(N, n_1+n_2)=SU(N)/SU(N-n_1-n_2)\times U(n_1+n_2)$ and
the flag manifold ${\cal M}^\prime= SU(N)/SU(N-n_1-n_2) \times U(n_1)\times
U(n_2)$, the NLSM describes the reduction of $U(n_1+n_2)$ gauge symmetry into
$U(n_1)\times U(n_2)$. This type of reduction and radiative mass
generations may provide  an alternative approach to the Higgs mechanism 
in the theories beyond the standard model or in the effective
field theory of QCD in the context of the hidden local symmetry
\cite{band}.

Finally, we mention possible relevance of our work with string theory.
We recall that the gauge symmetry
enhancement \cite{duff} and target space duality \cite{give} in string
theory have attracted an extensive study recently. Target space in the
large-$N$ limit could be unrealistic as space-time. Nevertheless, our results 
could provide us with some insight to study these subjects for strings moving
on curved backgrounds.

\vspace{5mm}


T.I. was supported by the grant of Post-Doc.~Program, Kyungpook 
National University (2000). 
P.O. was supported by the KOSEF through Project 
No. 2000-1-11200-001-3 and in part by the BK21 Physics Research Program.



\begin{references}

\bibitem{band} M. Bando, T. Kugo, and K. Yamawaki, Phys. Rep. {\bf 164}, 217
(1988). 

\bibitem{zakr} W. J. Zakrzewski, {\it Low Dimensional Sigma Models}
(IOP, Bristol, 1989).

\bibitem{poly} S. Coleman, {\it Aspects of Symmetry}
(Cambridge University Press, Cambridge, England, 1985);
A.M. Polyakov, {\it Gauge Fields and Strings} 
(Harwood, Chur, Switzerland, 1987). 

\bibitem{bard} W. A. Bardeen, B. W. Lee, and R. E. Shrock,
Phys. Rev. D {\bf 14}, 985 (1976). 

\bibitem{aref} I. Ya. Aref'eva and S.I. Azakov, Nucl. Phys. {\bf B162},
298 (1980); I. Ya. Aref'eva, Ann. Phys. (N.Y.) {\bf 117}, 393 (1979).

\bibitem{itoh} T. Itoh and P. Oh, Phys. Lett. B {\bf 491}, 362 (2000);
Phys. Rev. D {\bf 63}, 025019 (2001).

\bibitem{golo} H. Eichenherr, Nucl. Phys. {\bf B146}, 215 (1978); 
V. Golo and A. Perelomov, Phys. Lett. {\bf 79B}, 112 (1978);
A. D'Adda, P. Di Vecchia, and M. L\"uscher, Nucl. Phys. {\bf B146},
63 (1978); E. Witten, {\it ibid.} {\bf B149}, 285 (1979).

\bibitem{pisa} R. D. Pisarski, Phys. Rev. D {\bf 20}, 3358 (1979);
E. Brezin, S. Hikami, and J. Zinn-Justin, Nucl. Phys. {\bf B165},
528 (1980).

\bibitem{helg} S. Helgason, {\it Differential Geometry, Lie Groups, and
Symmetric Spaces} (Academic, New York, 1978).

\bibitem{duer} E. Gava, R. Jengo, and C. Omero,
Nucl. Phys. {\bf B158}, 381 (1979); S. Duane,
{\it ibid.} {\bf B168}, 32 (1980); G. Duerksen, Phys. Rev. D 
{\bf 24}, 926 (1981); M. Bando, Y. Taniguchi, and S. Tanimura,
Prog. Theor. Phys. {\bf  97}, 665 (1997). 

\bibitem{macf} This Lagrangian was considered for $r=1$ in 
A. J. Macfarlane, Phys. Lett. {\bf 82B}, 239 (1979). For our purposes, 
it is essential to keep $r$ arbitrary from the beginning. 

\bibitem{bal} E. Cremmer and B. Julia, 
Phys. Lett. {\bf 80B}, 48 (1978); Nucl. Phys. {\bf B159}, 141 (1979);
A. P. Balachandran, A. Stern, and C.G. Trahern, Phys. Rev. D {\bf 19}, 2416
(1979); M. Bando, T. Kugo, and K. Yamawaki, Prog. Theor. Phys. {\bf 73},
1541 (1985).

\bibitem{itoh1} T. Itoh, P. Oh, and C. Ryou, hep-th/0104204.

\bibitem{break} We stress that the nomenclature ``symmetry reduction'' here 
stands for the gauge symmetry away from the self-dual points being smaller 
than that along these points with $r=1$. We simply compare the gauge 
symmetries for different values of $r$ each of which corresponds to a 
different theory. This is not the usual dynamical Higgs mechanism in which 
we compare different phases for a fixed $r$.

\bibitem{jack} See R. Jackiw, Int. J. Mod. Phys. B {\bf 14}, 2011 (2000), 
and references therein.

\bibitem{duff} M. B. Green, J. H. Schwarz, and E. Witten, 
{\it Superstring Theory} 
(Cambridge University Press, Cambridge, England, 1987), Vols. I and II; 
J. Polchinski, {\it String theory} 
(Cambridge University Press, Cambridge, 1998), Vols. I and II.

\bibitem{give} A. Giveon, M. Porrati, and E. Rabinovici,
Phys. Rep. {\bf 244}, 77 (1994).

\end{references}
\end{document}